\documentstyle[prd,aps]{revtex}
\begin{document}
%
%
\draft
\title{On the use of the reciprocal basis in neutral meson mixing}
\author{Jo\~ao P.\ Silva\footnote{Permanent address: 
	Instituto Superior de Engenharia de Lisboa,
	Rua Conselheiro Em\'{\i}dio Navarro,
	1900 Lisboa, Portugal.}}
\address{Stanford Linear Accelerator Center, Stanford University, Stanford,
	 California 94309}
\date{\today}
\maketitle
\begin{abstract}
In the presence of CP violation,
the effective Hamiltonian matrix describing a neutral meson anti-meson
system does not commute with its hermitian conjugate.
As a result,
this matrix cannot be diagonalized by a unitary transformation
and one needs to introduce a reciprocal basis.
Although known,
this fact is seldom discussed and almost never used.
Here,
we use this concept to highlight a parametrization of the Hamiltonian
matrix in terms of physical observables,
and we show that using it reduces a number of long and tedious
derivations into simple matrix multiplications.
These results have a straightforward application for
propagation in matter.
We also comment on the (mathematical) relation with neutrino oscillations.
\end{abstract}


\section{Introduction}

We are interested in the effective $2 \times 2$ Hamiltonian matrix
describing the mixing in the $P^0 - \overline{P^0}$ systems,
where $P$ stands for $K$, $D$, $B_d$, or $B_s$.
We denote this $2 \times 2$ matrix by 
$\mbox{\boldmath $H$} = \mbox{\boldmath $M$} -i/2 \mbox{\boldmath $\Gamma$}$
where
\begin{equation}
\mbox{\boldmath $M$} = 
\left( \mbox{\boldmath $H$} + \mbox{\boldmath $H$}^\dagger \right)/2
\ \ \mbox{and}\ \ \
-i \mbox{\boldmath $\Gamma$}/2 = 
\left( \mbox{\boldmath $H$} - \mbox{\boldmath $H$}^\dagger \right)/2,
\end{equation}
describe the hermitian and anti-hermitian
parts of $\mbox{\boldmath $H$}$,
respectively.
Both $\mbox{\boldmath $M$}$ and $\mbox{\boldmath $\Gamma$}$ are hermitian.
Matrices satisfying 
$[ \mbox{\boldmath $H$}, \mbox{\boldmath $H$}^\dagger] = 0$
are called ``normal'' matrices.
It is easy to show that
\begin{equation}
\left[ \mbox{\boldmath $H$}, \mbox{\boldmath $H$}^\dagger \right] = 0 
\Leftrightarrow 
\left[ \mbox{\boldmath $M$}, \mbox{\boldmath $\Gamma$} \right] = 0.
\end{equation}
Moreover,
a matrix is normal if and only if it can be diagonalized by a
unitary transformation.
It is often stated that non-unitary transformations arise whenever
$\mbox{\boldmath $H$}$ is not hermitian.
This is not the case.
What is relevant is whether $\mbox{\boldmath $H$}$ is normal or not.
Indeed,
if $\mbox{\boldmath $\Gamma$} \neq 0$ then 
$\mbox{\boldmath $H$}$ is not hermitian;
however,
$\mbox{\boldmath $H$}$ can still be diagonalized by a unitary matrix as long
as $\left[ \mbox{\boldmath $M$}, \mbox{\boldmath $\Gamma$} \right] = 0$.

In section \ref{sec:reciprocal} we introduce the concept of
`reciprocal basis' and we show that the presence of T violation
in the $P^0 - \overline{P^0}$ system forces us to use such a basis.
The physical observables are defined in section \ref{sec:observables}
and they are used in section \ref{sec:relations} to parametrize
$\mbox{\boldmath $H$}$ exclusively in terms of measurable quantities.
The time evolution of the $P^0 - \overline{P^0}$ system is
discussed in section \ref{sec:time-evolution}.
Section \ref{sec:cascade} explains why the $P^0 - \overline{P^0}$
should be considered as intermediate states,
and section \ref{sec:BJpsiKS} shows an error which arises when
one does not use the reciprocal basis.
Matter effects are then considered in
section \ref{sec:matter}.
This differs from all previous analyses of matter effects in that no
use is made  of the Good equations;
here the time evolution is obtained in a trivial way.
In section~\ref{sec:boost} we compare the mixing in the
$P^0 - \overline{P^0}$ system with the mixing in the neutrino sector.
To this end,
we start by showing how the equation describing the
time evolution and its solution would look if we had chosen a
different reference frame.
We present our conclusions in section~\ref{sec:conclusions}.
For completeness,
appendix~\ref{app:A} contains some elementary notions of collision
theory,
which are needed to describe the evolution in the presence
of interactions with matter.
Appendix~\ref{app:others} contains two other parametrizations of
the physical observables commonly found in the literature,
the first of which is most convenient for the
comparison with the neutrino sector.

\section{The reciprocal basis}
\label{sec:reciprocal}

\subsection{Definition}

Why do we change basis at all?
One reason is that the time evolution of the state 
$|\psi(t)\rangle$ describing the $P^0 - \overline{P^0}$
mixed state,
which is given by
\begin{equation}
i \frac{d}{dt} |\psi(t)\rangle = \mbox{\boldmath $H$} |\psi(t)\rangle,
\label{Schrodinger-equation}
\end{equation}
becomes trivial in the basis in which $\mbox{\boldmath $H$}$ is diagonal.
Eq.~(\ref{Schrodinger-equation}) and 
$\mbox{\boldmath $H$}$ have been written in the
$P^0 - \overline{P^0}$ rest frame and $t$ is the proper time.

We denote the (complex) eigenvalues of $\mbox{\boldmath $H$}$ by 
$\mu_a=m_a -i/2 \Gamma_a$ and
$\mu_b=m_b -i/2 \Gamma_b$,
corresponding to the eigenvectors 
\begin{eqnarray}
\left(
\begin{array}{c}
| P_a \rangle \\ | P_b \rangle
\end{array}
\right)
=
\left(
\begin{array}{cc}
p_a & q_a\\ p_b & - q_b
\end{array}
\right)
\
\left(
\begin{array}{c}
| P^0 \rangle \\ | \overline{P^0} \rangle
\end{array}
\right)
= 
\mbox{\boldmath $X$}^T\ 
\left(
\begin{array}{c}
| P^0 \rangle \\ | \overline{P^0} \rangle
\end{array}
\right).
\label{PaPb}
\end{eqnarray}
As a result,
the matrix $\mbox{\boldmath $H$}$ is diagonalized through
\begin{equation}
\mbox{\boldmath $X$}^{-1} \mbox{\boldmath $H$} \mbox{\boldmath $X$} = 
\left(
\begin{array}{cc}
\mu_a & 0\\ 0 & \mu_b
\end{array}
\right),
\label{diagonalization}
\end{equation}
where
\begin{equation}
\mbox{\boldmath $X$}^{-1} = \frac{1}{p_a q_b+p_b q_a}
\left(
\begin{array}{cc}
q_b & p_b\\ q_a & - p_a
\end{array}
\right).
\end{equation}
As stated above, $\mbox{\boldmath $H$}$ is normal if and only if 
$\mbox{\boldmath $X$}$ is unitary.
This is what one learns in algebra.

So, why do (most) people worry about performing non-unitary transformations?
The reason is that one would like the mass basis
$\{| P_a \rangle, | P_b \rangle \}$ 
to retain a number of the nice (orthogonality) features of the
$\{| P^0 \rangle, | \overline{P^0} \rangle \}$ flavor
basis.
Among these:
the orthogonality conditions
\begin{eqnarray}
\langle P^0 | \overline{P^0} \rangle =
\langle \overline{P^0} | P^0  \rangle 
&=& 0,
\nonumber\\
\langle P^0 | P^0 \rangle =
\langle \overline{P^0} | \overline{P^0} \rangle &=& 1;
\end{eqnarray}
the fact that $| P^0 \rangle \langle P^0|$ and
$|\overline{P^0} \rangle \langle \overline{P^0}|$
are projection operators;
the completeness relation
\begin{equation}
| P^0 \rangle \langle P^0|
+
|\overline{P^0} \rangle \langle \overline{P^0}|
=
1;
\end{equation}
and the decomposition of the effective Hamiltonian as
\begin{eqnarray}
{\cal H} 
&=&
| P^0 \rangle H_{11} \langle P^0|
+
| P^0 \rangle H_{12} \langle \overline{P^0}|
+
| \overline{P^0} \rangle H_{21} \langle P^0|
+
| \overline{P^0} \rangle H_{22} \langle \overline{P^0}|
\nonumber\\*[3mm]
&=&
\left(
\begin{array}{cc}
| P^0 \rangle, & | \overline{P^0} \rangle
\end{array}
\right)
\mbox{\boldmath $H$}
\left(
\begin{array}{c}
\langle P^0 | \\
\langle \overline{P^0}|
\end{array}
\right).
\label{decomposition}
\end{eqnarray}
All these relations involve the basis of flavor eigenkets
$\{| P^0 \rangle, | \overline{P^0} \rangle \}$
and the basis of the corresponding bras
$\{\langle P^0 |, \langle \overline{P^0}| \}$.
The problem is that,
when $\mbox{\boldmath $H$}$ is not normal,
we {\em cannot} find similar relations involving the basis of mass eigenkets
$\left\{| P_a \rangle, | P_b \rangle \right\}$
and the basis of the corresponding bras,
$\left\{ \langle P_a |, \langle P_b | \right\}$.
In particular,
it is easy to see from the diagonalization in
Eq.~(\ref{diagonalization}) that the analogue of Eq.~(\ref{decomposition})
is
\begin{eqnarray}
{\cal H}
&=&
| P_a \rangle \mu_a \langle \tilde P_a|
+
| P_b \rangle \mu_b \langle \tilde P_b|
\nonumber\\*[3mm]
&=&
\left(
\begin{array}{cc}
| P_a \rangle, & | P_b \rangle
\end{array}
\right)
\left(
\begin{array}{cc}
\mu_a & 0\\
0 & \mu_b
\end{array}
\right)
\left(
\begin{array}{c}
\langle \tilde P_a | \\
\langle \tilde P_b |
\end{array}
\right).
\label{decomposition-mass}
\end{eqnarray}
This does not involve the bras $\langle P_a |$ and $\langle P_b|$,
\begin{equation}
\left(
\begin{array}{c}
\langle  P_a | \\ \langle  P_b | 
\end{array}
\right)
=
\mbox{\boldmath $X$}^{\dagger}
\left(
\begin{array}{c}
\langle P^0 | \\ \langle \overline{P^0} |
\end{array}
\right),
\label{wrong-basis}
\end{equation}
but rather the so called `reciprocal basis'
\begin{equation}
\left(
\begin{array}{c}
\langle \tilde P_a | \\ \langle \tilde P_b | 
\end{array}
\right)
=
\mbox{\boldmath $X$}^{-1}
\left(
\begin{array}{c}
\langle P^0 | \\ \langle \overline{P^0} |
\end{array}
\right).
\label{PtildeaPtildeb}
\end{equation}
The reciprocal basis may also be defined by the orthogonality
conditions
\begin{eqnarray}
\langle \tilde P_a | P_b \rangle = \langle \tilde P_b | P_a \rangle &=& 0,
\nonumber\\
\langle \tilde P_a | P_a \rangle = \langle \tilde P_b | P_b \rangle &=& 1.
\label{definition-reciprocal}
\end{eqnarray}
Moreover,
$|P_a \rangle \langle \tilde P_a|$ and
$|P_b \rangle \langle \tilde P_b|$ are projection operators,
and  the partition of unity becomes
\begin{equation}
|P_a \rangle \langle \tilde P_a| + |P_b \rangle \langle \tilde P_b|=1.
\label{partition-of-unity}
\end{equation}
If $\mbox{\boldmath $H$}$ is not normal,
then $\mbox{\boldmath $X$}$ is not unitary,
and
$\left\{ \langle P_a |, \langle P_b | \right\}$
in Eq.~(\ref{wrong-basis}) do not coincide with
$\left\{ \langle \tilde P_a |, \langle \tilde P_b | \right\}$
in Eq.~(\ref{PtildeaPtildeb}).
Another way to state this fact is to note that 
$\mbox{\boldmath $H$}$ is normal
($\mbox{\boldmath $X$}$ is unitary)
if and only if its right-eigenvectors 
coincide with its left-eigenvectors.

That these features have an impact on the $K^0 - \overline{K^0}$
system,
was pointed out long ago by Sachs \cite{Sac63,Sac64} and by Enz and
Lewis \cite{Enz65}.
More recently,
they have been stressed by Alvarez-Gaum\'{e} {\it et al.} \cite{Alv99},
and by Branco, Lavoura and Silva in their book ``CP violation'' \cite{BLS}.
Still,
we have found that they are not common knowledge.
This is unfortunate since there are a number of results that usually
require considerable algebra which become trivial once the
matrix formulation discussed here is implemented.
Moreover,
one can express the matrix elements of $\mbox{\boldmath $H$}$,
written in the $P^0 -\overline{P^0}$ basis,
in terms of observable quantities.
This is what we show here.

\subsection{The relation to CP violation}
\label{sec:CPT}

We will now show that the reciprocal basis is required by
the observation of T and CP violation in the mixing in the neutral
meson systems.
The discrete symmetries have the following effects on the
matrix elements of $\mbox{\boldmath $H$}$:
\begin{eqnarray}
{\cal CPT}\ \mbox{conservation}
& \Rightarrow & H_{11} = H_{22},
\nonumber\\ 
{\cal T}\ \mbox{conservation}
& \Rightarrow & |H_{12}| = |H_{21}|,
\nonumber\\ 
{\cal CP}\ \mbox{conservation}
& \Rightarrow & H_{11} = H_{22}\ 
\mbox{and}\ |H_{12}| = |H_{21}|.
\end{eqnarray}
The 1964 discovery that $|H_{12}| \neq |H_{21}|$ in the kaon system
\cite{Chr64} means that there is T and CP violation in
$K^0 - \overline{K^0}$ mixing.
Moreover,
since the $(1,1)$ entry in the matrix 
$[\mbox{\boldmath $H$} , \mbox{\boldmath $H$}^\dagger]$
is given by $|H_{12}|^2 - |H_{21}|^2$,
this experimental result also implies that the matrix $\mbox{\boldmath $H$}$
is not normal and,
thus,
that we are forced to deal with non-unitary matrices in the neutral kaon
system.

For the other neutral meson systems,
$|H_{12}| \neq |H_{21}|$ has not been established experimentally.
Nevertheless,
the Standard Model predicts that,
albeit the difference is small,
$|H_{12}| \neq |H_{21}|$ does indeed hold.
As before,
this implies CP violation in the mixing and forces the use of 
the reciprocal basis in all the neutral meson systems.

\section{Observables in the $P^0 - \overline{P^0}$ mixing}
\label{sec:observables}

Let us start by introducing some notation.
We define
\begin{eqnarray}
\mu = m - i \Gamma/2 &\equiv& (\mu_a + \mu_b)/2,
\nonumber\\
\Delta \mu = \Delta m - i \Delta \Gamma/2 &\equiv& \mu_a - \mu_b.
\end{eqnarray}
Sometimes it is convenient to trade the eigenvalue difference for
$x - i y \equiv \Delta \mu / \Gamma$.
We may write the mixing matrix 
$\mbox{\boldmath $X$}$ 
in terms of new parameters
\begin{equation}
\theta =
\frac{\frac{q_a}{p_a} - \frac{q_b}{p_b}}{
\frac{q_a}{p_a} + \frac{q_b}{p_b}},
\label{theta}
\end{equation}
and
\begin{equation}
\frac{q}{p} = \sqrt{\frac{q_a q_b}{p_a p_b}}
\label{ratio}
\end{equation}
Notice that we have not defined the quantities $q$ and $p$ separately;
we only define the ratio $q/p$.
With this notation the mixing matrix may be re-written as
\begin{equation}
\mbox{\boldmath $X$} =
\left(
\begin{array}{cc}
1 & 1\\
\frac{q}{p} \sqrt{\frac{1+\theta}{1-\theta}} &
- \frac{q}{p} \sqrt{\frac{1-\theta}{1+\theta}}
\end{array}
\right)
\left(
\begin{array}{cc}
p_a & 0\\
0 & p_b
\end{array}
\right),
\label{X-paramet}
\end{equation}
\begin{equation}
\mbox{\boldmath $X$}^{-1} =
\left(
\begin{array}{cc}
p_a^{-1} & 0\\
0 & p_b^{-1}
\end{array}
\right)
\left(
\begin{array}{cc}
\frac{1-\theta}{2} & \frac{p}{q} \frac{\sqrt{1-\theta^2}}{2}\\
\frac{1+\theta}{2} & - \frac{p}{q} \frac{\sqrt{1-\theta^2}}{2}
\end{array}
\right).
\label{X-1-paramet}
\end{equation}
We point out that these transformation matrices involve
the normalization constants $p_a$ and $p_b$.
Finally,
it will also prove convenient to define
\begin{equation}
\delta
=
\frac{1 - \left| \frac{q}{p} \right|^2}{
1 + \left| \frac{q}{p} \right|^2},
\end{equation}
meaning that $|q/p| = \sqrt{\frac{1-\delta}{1+\delta}}$.

The fact that the trace and determinant are invariant under the general
similarity transformation in Eq.~(\ref{diagonalization}) implies that
\begin{eqnarray}
\mu &=& (H_{11}+H_{22})/2,
\nonumber\\
\Delta \mu &=& \sqrt{4 H_{12} H_{21} + (H_{22} - H_{11})^2}.
\label{eigenvalues}
\end{eqnarray}
Moreover,
from
\begin{eqnarray}
\left( \begin{array}{cc} H_{11} & H_{12} \\ H_{21} & H_{22} \end{array} 
\right)
\left( \begin{array}{c} p_a \\ q_a \end{array} \right)
&=& \mu_a
\left( \begin{array}{c} p_a \\ q_a \end{array} \right),
\nonumber\\
\left( \begin{array}{cc} H_{11} & H_{12} \\ H_{21} & H_{22} \end{array} 
\right)
\left( \begin{array}{c} p_b \\ - q_b \end{array} \right)
&=& \mu_b
\left( \begin{array}{c} p_b \\ - q_b \end{array} \right).
\label{eigenvector-equations}
\end{eqnarray}
we find that
\begin{eqnarray}
\frac{q_a}{p_a}
&=& \frac{\mu_a - H_{11}}{H_{12}}
= \frac{H_{21}}{\mu_a - H_{22}},
\nonumber\\
\frac{q_b}{p_b}
&=& \frac{H_{11} - \mu_b}{H_{12}}
= \frac{H_{21}}{H_{22} - \mu_b},
\label{eigenvectors}
\end{eqnarray}
leading to
\begin{eqnarray}
\theta &=&
\frac{H_{22} - H_{11}}{\mu_a - \mu_b},
\nonumber\\
\delta &=&
\frac{|H_{12}|-|H_{21}|}{|H_{12}|+|H_{21}|},
\label{mixing-observables}
\end{eqnarray}
and $q/p = \sqrt{H_{21}/H_{12}}$.
We see that $\mbox{Re}\, \theta$ and $\mbox{Im}\, \theta$ are
CP- and CPT-violating,
while $\delta$ is CP- and T-violating.

Although $\mbox{\boldmath $H$}$ contains eight real numbers,
only seven are physically meaningful.
Indeed,
one is free to change the phase
of the kets $| P^0 \rangle$,
$| \overline{P^0} \rangle$,
$| P_a \rangle$,
and $| P_b \rangle$,
as
\begin{eqnarray}
| P^0 \rangle & \rightarrow & e^{i \gamma} | P^0 \rangle,
\nonumber\\
| \overline{P^0} \rangle & \rightarrow & e^{i \overline{\gamma}}
| \overline{P^0} \rangle,
\nonumber\\
| P_a \rangle & \rightarrow & e^{i \gamma_a} | P_a \rangle,
\nonumber\\
| P_b \rangle & \rightarrow & e^{i \gamma_b} | P_b \rangle.
\label{ket rephasing}
\end{eqnarray}
Under these transformations
\begin{eqnarray}
H_{12} & \rightarrow & e^{i \left( \overline{\gamma} - \gamma \right)} H_{12},
\nonumber\\
H_{21} & \rightarrow & e^{i \left( \gamma - \overline{\gamma} \right)} H_{21},
\nonumber\\
q/p & \rightarrow & e^{i \left( \gamma - \overline{\gamma} \right)} q/p,
\label{rephasing}
\end{eqnarray}
while $H_{11}$, $H_{22}$, $\mu$, $\Delta \mu$,
$\theta$, and $\delta$ do not change.
Therefore,
the relative phase between $H_{12}$ and $H_{21}$ is physically
meaningless and $\mbox{\boldmath $H$}$ contains only seven observables.
Similarly,
{\em the phase of $q/p$ is also unphysical}.
As a result,
we have four observables in the eigenvalues,
$\mu$ and $\Delta \mu$,
and three in the mixing matrix,
$\theta$ and $\delta$ (or, alternatively, $|q/p|$).

\section{Parametrizing $\mbox{\boldmath $H$}$ with measurable quantities}
\label{sec:relations}

Eqs.~(\ref{eigenvalues}) and (\ref{mixing-observables})
give the measurable mixing and eigenvalue parameters in
terms of the $H_{ij}$ matrix elements which one can calculate
in a given model.
Given the current and upcoming experimental probes of the
various neutral meson systems,
it seems much more appropriate to do precisely the opposite;
that is, to give the $H_{ij}$ matrix elements in terms of the
experimentally accessible quantities.
Such expressions would give $M_{ij}$ and $\Gamma_{ij}$ in
a completely model independent way,
with absolutely no assumptions.
One could then calculate these quantities in any given model;
if they fit in the allowed ranges the model would be viable.

Surprisingly,
this is not is done in most expositions of the
$P^0 - \overline{P^0}$ mixing.
The reason is simple.
Eqs.~(\ref{eigenvalues}) and (\ref{mixing-observables})
are non-linear in the $H_{ij}$ matrix elements.
Thus,
inverting them by brute force would entail a tedious calculation.
With the matrix manipulation discussed here this inversion
is straightforward.
Indeed,
Eq.~(\ref{diagonalization}) can be trivially transformed into
\cite{before}
\begin{eqnarray}
\mbox{\boldmath $H$} &=&
\mbox{\boldmath $X$}
\left(
\begin{array}{cc}
\mu_a & 0\\
0 & \mu_b
\end{array}
\right)
\mbox{\boldmath $X$}^{-1}
\nonumber\\
&=&
\left(
\begin{array}{cc}
\mu - \frac{\Delta \mu}{2} \theta \ &
\frac{p}{q} \frac{\sqrt{1-\theta^2}}{2} \Delta \mu
\\*[2mm]
\frac{q}{p} \frac{\sqrt{1-\theta^2}}{2} \Delta \mu \ &
\mu + \frac{\Delta \mu}{2} \theta
\end{array}
\right),
\label{master}
\end{eqnarray}
where we have used Eqs.~(\ref{X-paramet}) and (\ref{X-1-paramet}).
This equation expresses in a very compact form 
the relation between the quantities which are experimentally accessible
and those which are easily calculated in a given theory.
Expanding it,
we find
\begin{eqnarray}
M_{11}
& = &
m - \mbox{Re}\, \theta\, \frac{\Delta m}{2} - 
\mbox{Im}\, \theta\, \frac{\Delta \Gamma}{4},
\nonumber\\
M_{22}
& = &
m + \mbox{Re}\, \theta\, \frac{\Delta m}{2} + 
\mbox{Im}\, \theta\, \frac{\Delta \Gamma}{4},
\nonumber\\
\frac{q}{p} M_{12} &=&
\frac{1}{2(1+\delta)}
\left[
\mbox{Re}(\sqrt{1-\theta^2})
\left( \Delta m - i \delta\, \frac{\Delta \Gamma}{2}\right)
+
\mbox{Im}(\sqrt{1-\theta^2})
\left( \frac{\Delta \Gamma}{2} + i \delta\, \Delta m \right)
\right],
\label{Mij-physical}
\end{eqnarray}
and
\begin{eqnarray}
\Gamma_{11}
& = &
\Gamma - \mbox{Re}\, \theta\, \frac{\Delta \Gamma}{2} + 
\mbox{Im}\, \theta\, \Delta m,
\nonumber\\
\Gamma_{22}
& = &
\Gamma + \mbox{Re}\, \theta\, \frac{\Delta \Gamma}{2} - 
\mbox{Im}\, \theta\, \Delta m,
\nonumber\\
\frac{q}{p} \Gamma_{12}
& = &
\frac{1}{1+\delta}
\left[
\mbox{Re}(\sqrt{1-\theta^2})
\left( \frac{\Delta \Gamma}{2} + i \delta\, \Delta m \right)
-
\mbox{Im}(\sqrt{1-\theta^2})
\left( \Delta m - i \delta\, \frac{\Delta \Gamma}{2}\right)
\right].
\label{Gammaij-physical}
\end{eqnarray}
We would argue that this is the best way to quote the
experimental results.
The impact of any assumption made about the physical observables,
such as CPT or T conservation,
is transparent in Eqs.~(\ref{Mij-physical}) and (\ref{Gammaij-physical}).

A few remarks are in order.
Firstly we note that Eqs.~(\ref{X-paramet}) and (\ref{X-1-paramet})
involved the overall normalization factors $p_a$ and $p_b$,
but that these cancel in the multiplication on the right hand side
of Eq.~(\ref{master}).
Secondly,
although $M_{12}$, $\Gamma_{12}$ and $q/p$ are not 
rephasing invariant,
we can see from Eqs.~(\ref{rephasing}) that $q/p\, M_{12}$,
$q/p\, \Gamma_{12}$ and $M_{12}\, \Gamma_{12}^\ast$
are indeed physically meaningful.
Thirdly,
the equations involving $\Gamma$ are needed also for the
unitarity conditions \cite{Lav99}
\begin{eqnarray}
\sum_g \left| A_g \right|^2 &=& 
\Gamma_{11}
=
\Gamma
\left( 1 - y\, \mbox{Re}\, \theta + x\, \mbox{Im}\, \theta \right),
\nonumber\\
\sum_g \left| \bar A_g \right|^2 &=& 
\Gamma_{22}
=
\Gamma
\left( 1 + y\, \mbox{Re}\, \theta - x\, \mbox{Im}\, \theta \right),
\nonumber\\
\sum_g \frac{q}{p} A_g^\ast \bar A_g &=& 
\frac{q}{p} \Gamma_{12}
=
\Gamma
\frac{\left( y + i \delta x \right) \mbox{Re} (\sqrt{1 - \theta^2})
- \left( x - i \delta y \right) \mbox{Im} (\sqrt{1 - \theta^2})}{1 + \delta},
\label{unitarity}
\end{eqnarray}
where $A_g = \langle g | T | P^0 \rangle$,
$\bar A_g = \langle g | T | \overline{P^0} \rangle$,
and the sums run over all the available decay modes $g$.

\section{Time evolution}
\label{sec:time-evolution}

The time evolution of the neutral meson system is easily obtained
using Eqs.~(\ref{decomposition-mass}) and (\ref{partition-of-unity}),
and the fact that $|P_a \rangle \langle \tilde P_a|$ and
$|P_b \rangle \langle \tilde P_b|$ are projection operators.
We find,
\begin{eqnarray}
\exp{(- i {\cal H} t)}
&=&
e^{-i \mu_a t} |P_a \rangle \langle \tilde P_a|
+
e^{-i \mu_b t} |P_b \rangle \langle \tilde P_b|.
\nonumber\\*[3mm]
&=&
\left(
\begin{array}{cc}
| P_a \rangle, & | P_b \rangle
\end{array}
\right)
\left(
\begin{array}{cc}
e^{-i \mu_a t} & 0\\
0 & e^{-i \mu_b t}
\end{array}
\right)
\left(
\begin{array}{c}
\langle \tilde P_a | \\
\langle \tilde P_b |
\end{array}
\right).
\label{ev-operator}
\end{eqnarray}
It is now trivial to write the evolution operator back in the
flavor basis.
Indeed,
using Eqs.~(\ref{PaPb}) and (\ref{PtildeaPtildeb}),
we find
\begin{eqnarray}
\exp{(- i {\cal H} t)}
&=&
\left(
\begin{array}{cc}
| P^0 \rangle, & | \overline{P^0} \rangle
\end{array}
\right)
\mbox{\boldmath $X$}
\left(
\begin{array}{cc}
e^{-i \mu_a t} & 0\\
0 & e^{-i \mu_b t}
\end{array}
\right)
\mbox{\boldmath $X$}^{-1}
\left(
\begin{array}{c}
\langle P^0 | \\
\langle \overline{P^0} |
\end{array}
\right)
\nonumber\\*[3mm]
&=&
\left(
\begin{array}{cc}
| P^0 \rangle, & | \overline{P^0} \rangle
\end{array}
\right)
\left(
\begin{array}{cc}
g_+(t) - \theta\, g_-(t)\  & \frac{p}{q} \sqrt{1  - \theta^2} g_-(t)\\*[2mm]
\frac{q}{p} \sqrt{1  - \theta^2} g_-(t)\  & g_+(t) + \theta\, g_-(t)
\end{array}
\right)
\left(
\begin{array}{c}
\langle P^0 | \\
\langle \overline{P^0} |
\end{array}
\right),
\label{evolution-in-general}
\end{eqnarray}
where
\begin{equation}
g_{\pm}(t) \equiv 
\frac{1}{2} \left( e^{-i \mu_a t} \pm e^{-i \mu_b t} \right)
=
e^{-i m t}\, e^{- \Gamma t/2}
\left\{
\begin{array}{c}
\cos{(\frac{\Delta \mu\, t}{2})}\\*[2mm]
-i \sin{(\frac{\Delta \mu\, t}{2})}
\end{array}
\right.
.
\label{g+-}
\end{equation}
This corresponds to the usual expressions for the time evolution 
of a state which starts out as $P^0$ or $\overline{P^0}$
\begin{eqnarray}
| P^0(t) \rangle
= \exp{(- i {\cal H} t)} | P^0 \rangle
&=&
\left[ g_+(t) - \theta\, g_-(t) \right] | P^0 \rangle
+
\frac{q}{p} \sqrt{1  - \theta^2} g_-(t)\, | \overline{P^0} \rangle,
\nonumber\\*[3mm]
| \overline{P^0}(t) \rangle
= \exp{(- i {\cal H} t)} | \overline{P^0} \rangle
&=&
\frac{p}{q} \sqrt{1  - \theta^2} g_-(t) | P^0 \rangle
+
\left[ g_+(t) + \theta\, g_-(t) \right] | \overline{P^0} \rangle,
\label{usual-time-evolution}
\end{eqnarray}
respectively.
At this point it is important to emphasize the fact that,
in deriving this result,
no assumptions were made about the form of the original matrix 
$\mbox{\boldmath $H$}$.
This observation will become important once we consider
the evolution in matter.

\section{Neutral mesons as intermediate states}
\label{sec:cascade}

Because there is CP violation in $P^0 - \overline{P^0}$ mixing,
there is no selection rule allowing us to choose a final state
$f$ to which $P_a$ (or $P_b$) can decay while $P_b$ ($P_a$) cannot.
That is,
all calculations must involve the full transition chain \cite{Kay98}
\begin{equation}
i \rightarrow X \{ P_a, P_b \} \rightarrow X f,
\label{cascade-decays}
\end{equation}
with both neutral meson eigenstates as intermediate states,
in order to be {\em formally correct}.
Obviously,
one could ignore this problem.
Still,
as we show in section \ref{sec:BJpsiKS},
one will be lead into incorrect results if the reciprocal basis
is not used as the `out' bra.

Recently,
Amorim, Santos, and Silva \cite{Amo99} have highlighted a very important point
about the transition chain in Eq.~(\ref{cascade-decays}).
They showed that this evolution can be fully parametrized by
the usual quantities $\lambda_f$ and $\lambda_{\bar f}$,
describing the decays $\{ P^0, \overline{P^0} \} \rightarrow f, \bar f$,
supplemented by two new quantities $\xi_i$ and $\xi_{\bar i}$,
describing the production mechanism 
$i, \bar i \rightarrow \{ P^0, \overline{P^0} \}$.
(Although they applied these results only to the case in which
$i, \bar i \rightarrow \{ P^0, \overline{P^0} \}$
represents a decay,
their formalism is valid in all generality.)
The new quantities $\xi_i$ and $\xi_{\bar i}$ may entail
new sources of CP violation,
just like $\lambda_f$ and $\lambda_{\bar f}$ do.
They are absent from the decays
$B \rightarrow J/\psi K \rightarrow J/\psi [f]_K$
studied previously \cite{All} because,
in those cases,
the initial $B^0$ meson can only decay to one
of the kaon's flavor eigenstates.
However,
they are crucial for the decays
$B^\pm \rightarrow D + X^\pm \rightarrow [f]_D + X^\pm$ \cite{Mec98},
and, in general,
whenever the initial state $i$ can produce (or, in particular, decay into)
both flavor eigenstates of the intermediate neutral meson system,
$P^0$ and $\overline{P^0}$.

Let us consider the decay chain
$i \rightarrow X \{ P_a, P_b \} \rightarrow X f$.
The complete amplitude for this process involves
the amplitude for the initial decay into $X P_a$ or $X P_b$,
the time-evolution amplitude for this state,
given by Eq.~(\ref{ev-operator}),
and finally the amplitude for the decay into $X f$.
Suppressing the reference to $X$,
we find
\begin{equation}
A \left( i \rightarrow P_{a,b} \rightarrow f \right)
=
\langle f | T | P_a \rangle\, e^{- i\, \mu_a t}\,
\langle \tilde P_a | T | i \rangle
+
\langle f | T | P_b \rangle\, e^{- i\, \mu_b t}\,
\langle \tilde P_b | T | i \rangle.
\label{vodoo-1}
\end{equation}
This is an exact expression.
However,
sometimes it is possible to choose a final state $f$
and to set the experimental conditions in such a way as
to maximize the importance of $ i \rightarrow X P_a \rightarrow X f$ 
relative to $ i \rightarrow X P_b \rightarrow X f$.
In that case we may make the approximation
\begin{eqnarray}
A \left( i \rightarrow P_{a,b} \rightarrow f \right)
&\approx&
A \left( i \rightarrow P_a \rightarrow f \right)
\nonumber\\
& = &
\langle f | T | P_a \rangle\, e^{- i\, \mu_a t}\,
\langle \tilde P_a | T | i \rangle
\nonumber\\
&=&
\langle f | T | P_a \rangle\, e^{- i\, \mu_a t}\,
\left[
\langle \tilde P_a | P^0 \rangle \langle P^0 | T | i \rangle
+
\langle \tilde P_a | \overline{P^0} \rangle
\langle \overline{P^0} | T | i \rangle
\right],
\nonumber\\
& &
\label{vodoo-2}
\end{eqnarray}
where we have used the partition of unity
$| P^0 \rangle \langle P^0 | + 
| \overline{P^0} \rangle \langle \overline{P^0} | = 1$
to derive the last line.
When one uses the approximation in Eq.~(\ref{vodoo-2}),
one talks about `the decay $i \rightarrow X P_a$',\footnote{
Nevertheless,
strictly speaking,
it is Eq.~(\ref{vodoo-1}) which expresses the correct way to think
about decays into neutral-meson eigenstates \cite{Enz65,Kay98}.
As we stressed above,
the point is that,
since CP is violated,
there is no final state $f$
that can be obtained
only from $P_a$ and not from $P_b$.
There will always be a non-zero amplitude
for the decay path $ i \rightarrow X P_b \rightarrow X f$.}
and writes
\begin{eqnarray}
A \left( i \rightarrow X P_a \right) &=&
\langle \tilde P_a | P^0\rangle\, A (i \rightarrow X P^0)
+ \langle \tilde P_a | \overline{P^0} \rangle\,
A (i \rightarrow X \overline{P^0})
\nonumber\\
&=&
\frac{1}{2} \left[ p^{-1} A (i \rightarrow X P^0)
+ q^{-1} A (i \rightarrow X \overline{P^0}) \right],
\label{vodoo-3}
\end{eqnarray}
where,
in the last line,
we have assumed the CPT-invariant case:
\begin{eqnarray}
\langle \tilde P_a | &=&
\frac{1}{2}
\left( p^{-1} \langle P^0 | + q^{-1} \langle \overline{P^0} | \right),
\nonumber\\*[3mm]
\langle \tilde P_b | &=&
\frac{1}{2}
\left( p^{-1} \langle P^0 | - q^{-1} \langle \overline{P^0} | \right).
\label{a subtileza essencial}
\end{eqnarray}
Therefore,
the ratio of the two component amplitudes
in Eq.~(\ref{vodoo-3}) is given by $q^{-1} / p^{-1} = p / q$,
and not by $q^\ast / p^\ast$---as would have been the case
if we had used $\langle P_H |$ instead of $\langle \tilde P_H |$.
The difference between $q^{-1} / p^{-1}$ and $q^\ast / p^\ast$
only disappears in the limit $\left| q/p \right| = 1$.
We will now show that this has a formal impact in the
study of the decay $B_d \rightarrow J/\psi K_S$.

\section{On the need for the reciprocal basis in $B_d \rightarrow J/\psi K_S$}
\label{sec:BJpsiKS}

This decay is so important that it is surprising how many times
it is calculated without even mentioning that the use of the
reciprocal basis is {\em required} in order to obtain the {\em exact} result.
We repeat,
in this decay the use of the reciprocal basis is not a convenient
calculational tool.
It is unavoidable when one wishes to obtain the result without approximations.

The first observation we should make is that what one looks for
experimentally is the decay chain
$B_d \rightarrow J/\psi K \rightarrow J/\psi (\pi \pi)_K$,
and that both intermediate $K_S$ and $K_L$ contribute to this decay.
The following argument should make it clear why
the intermediate $K_L$ must contribute.
Consider the decay chain
$B_d \rightarrow J/\psi K \rightarrow J/\psi (\pi \pi)_K$,
but where we have chosen to look only
for kaons which live a proper time $\tau \gg \tau_S$ before they decay.
Clearly,
for these kaons,
the $K_S$ component will have disappeared before the decay,
and  all $\pi \pi$ final states must have come from an intermediate $K_L$.
This explains why, in general, one must use Eq.~(\ref{vodoo-1}).
However,
in the experiments searching for $B_d \rightarrow J/\psi K_S$ one
is looking at kaon proper times $\tau \leq 10 \tau_S$.
Therefore,
in these experiments the decay path 
$B_d \rightarrow J/\psi K_L \rightarrow J/\psi (\pi \pi)_K$
is very suppressed with respect to the decay path
$B_d \rightarrow J/\psi K_S \rightarrow J/\psi (\pi \pi)_K$
both due to the huge ratio 
$A(K_S \rightarrow \pi \pi)/A(K_L \rightarrow \pi \pi)$
and to the time interval probed.
This leads us to Eq.~(\ref{vodoo-2}) and,
ignoring the normalizations $p_a$ and $p_b$,
allows us to talk about the decay $B_d \rightarrow J/\psi K_S$ as in
Eq.~(\ref{vodoo-3}).

Having established under which circumstances we may (to good approximation)
talk about the decay $B_d \rightarrow J/\psi K_S$,
we are now in position to describe the upcoming measurement
of CP violation in this decay.
These experiments will determine the imaginary part of
\begin{equation}
\lambda_{B_d \rightarrow J/\psi K_S}
\equiv
\frac{q_{Bd}}{p_{Bd}} \frac{A(\overline{B_d} \rightarrow J/\psi K_S)}{
A(B_d \rightarrow J/\psi K_S)}.
\end{equation}
We wish to calculate $A(B_d \rightarrow J/\psi K_S)$ and
$A(\overline{B_d} \rightarrow J/\psi K_S)$.
We recall that the decays $B_d \rightarrow J/\psi \overline{K^0}$
and $\overline{B_d} \rightarrow J/\psi K^0$ are forbidden to leading
order in the SM,
and, to simplify the problem,
we consider the CPT-conserving case,
in which
\begin{eqnarray}
| K_S \rangle &=& p_K |K^0 \rangle - q_K |\overline{K^0} \rangle,
\nonumber\\
\langle K_S | &=& p_K^\ast \langle K^0 | - q_K^\ast \langle \overline{K^0}|,
\nonumber\\
\langle \tilde K_S | &=& \frac{1}{2} \left[
p_K^{-1} \langle K^0 | - q_K^{-1} \langle \overline{K^0}| \right].
\end{eqnarray}
The question is whether one should use
$\langle \tilde K_S|$ or $\langle K_S|$ in the final state.
That is,
we wish to know whether to use
\begin{eqnarray}
A \left( B_d \rightarrow J/\psi K_S \right) &=&
\langle \tilde K_S | K^0\rangle\, A (B_d \rightarrow J/\psi K^0)
+ \langle \tilde K_S | \overline{K^0} \rangle\,
A (B_d \rightarrow J/\psi \overline{K^0})
\nonumber\\
&=&
\frac{1}{2} \left[ p_K^{-1} A (B_d \rightarrow J/\psi K^0)
- q_K^{-1} A (B_d \rightarrow J/\psi \overline{K^0}) \right]
\nonumber\\
&=&
\frac{1}{2} p_K^{-1} A (B_d \rightarrow J/\psi K^0),
\label{Bd-correct}
\end{eqnarray}
and
\begin{equation}
A \left( \overline{B_d} \rightarrow J/\psi K_S \right) =
- \frac{1}{2} q_K^{-1} A (\overline{B_d} \rightarrow J/\psi \overline{K^0})
\end{equation}
or, alternatively,
use
\begin{eqnarray}
A \left( B_d \rightarrow J/\psi K_S \right) &=&
\langle K_S | K^0\rangle\, A (B_d \rightarrow J/\psi K^0)
+ \langle  K_S | \overline{K^0} \rangle\,
A (B_d \rightarrow J/\psi \overline{K^0})
\nonumber\\
&=&
p_K^{\ast} A (B_d \rightarrow J/\psi K^0)
- q_K^{\ast} A (B_d \rightarrow J/\psi \overline{K^0})
\nonumber\\
&=&
p_K^{\ast} A (B_d \rightarrow J/\psi K^0),
\label{Bd-incorrect}
\end{eqnarray}
and
\begin{equation}
A \left( \overline{B_d} \rightarrow J/\psi K_S \right) =
- q_K^{\ast} A (\overline{B_d} \rightarrow J/\psi \overline{K^0}).
\end{equation}
In the first case we obtain
\begin{equation}
\lambda_{B_d \rightarrow J/\psi K_S}
\equiv
- \frac{q_{Bd}}{p_{Bd}}
\frac{A(\overline{B_d} \rightarrow J/\psi \overline{K^0})}{
A(B_d \rightarrow J/\psi K^0)}
\frac{p_K}{q_K},
\label{lambda-correct}
\end{equation}
in the second we obtain
\begin{equation}
\lambda_{B_d \rightarrow J/\psi K_S}
\equiv
- \frac{q_{Bd}}{p_{Bd}}
\frac{A(\overline{B_d} \rightarrow J/\psi \overline{K^0})}{
A(B_d \rightarrow J/\psi K^0)}
\frac{q^\ast_K}{p^\ast_K}
\label{lambda-incorrect}.
\end{equation}
From the previous section,
we know that the first expression is the correct one.
And,
in deriving it,
we had to know what the reciprocal basis was and that it had to be used.
Nevertheless,
since $|q_K/p_K|$ only differs from one at order $10^{-3}$ and we are
looking for a large effect in $\lambda_{B_d \rightarrow J/\psi K_S}$,
this detail,
although needed for an {\em exact formal} derivation,
is numerically insignificant.
This explains why it has gone largely unnoticed \cite{GKN}.

\section{Matter effects in the $P^0 - \overline{P^0}$ evolution}
\label{sec:matter}

We now wish to study how the time evolution of the $P^0 - \overline{P^0}$
changes in the presence of matter.
It should be clear that the matter effects will change
the specific form of $\mbox{\boldmath $H$}$ but,
since we have considered the most general such matrix,
all the derivations presented above should still apply.
It remains to relate the parameters in matter and in vacuum.

We will denote the matrices, matrix elements and eigenvalues
in vacuum by unprimed quantities and their analogue
in matter by primed quantities.
For example,
when kaons transverse matter,
they are subject to strong interactions which conserve strangeness
but which treat the $K^0$ and $\overline{K^0}$ 
differently.\footnote{The total cross-section for $\overline{K^0}$
interacting with a nucleus is larger than that for $K^0$ on the same nucleus.
For example,
$\overline{K^0} p \rightarrow \Lambda \pi^+$
takes place but there is no corresponding reaction for $K^0$.}
This effect may be parametrized by a new effective Hamiltonian
\begin{equation}
\mbox{\boldmath $H$}_{\rm nuc} =
\left(
\begin{array}{cc}
\chi & 0\\
0 & \bar \chi
\end{array}
\right),
\label{nuc}
\end{equation}
which must be added to the Hamiltonian in vacuum.
Notice that this parametrization is completely general.
It describes any strangeness-conserving interaction whatsoever.
It is also important to notice that our original evolution equation,
Eq.~(\ref{Schrodinger-equation}),
and vacuum Hamiltonian $\mbox{\boldmath $H$}$ have been written in the
$P^0 - \overline{P^0}$ rest frame.
Before we add $\mbox{\boldmath $H$}_{\rm nuc}$ to 
$\mbox{\boldmath $H$}$ we must ensure that
$\mbox{\boldmath $H$}_{\rm nuc}$ is also expressed in the rest frame.
This point is discussed in appendix~\ref{app:A}.

The full Hamiltonian in matter becomes
\begin{equation}
\mbox{\boldmath $H$}^\prime = 
\mbox{\boldmath $H$} + \mbox{\boldmath $H$}_{\rm nuc}.
\label{hh-in-matter}
\end{equation}
Now,
we have already studied the most general effective Hamiltonian,
and Eq.~(\ref{master}) relates such an Hamiltonian written in the
flavor basis with the corresponding eigenvalues and mixing parameters.
Therefore,
relating the observables in vacuum and in matter becomes another
simple exercise.
Eqs.~(\ref{master}), (\ref{nuc}) and (\ref{hh-in-matter}) yield
\begin{equation}
\left(
\begin{array}{cc}
\mu^\prime - \frac{\Delta \mu^\prime}{2} \theta^\prime \ &
\frac{p^\prime}{q^\prime} \frac{\sqrt{1-\theta^{\prime 2}}}{2}
\Delta \mu^\prime
\\*[2mm]
\frac{q^\prime}{p^\prime} \frac{\sqrt{1-\theta^{\prime 2}}}{2}
\Delta \mu^\prime \ &
\mu^\prime + \frac{\Delta \mu^\prime}{2} \theta^\prime
\end{array}
\right)
=
\left(
\begin{array}{cc}
\mu - \frac{\Delta \mu}{2} \theta \ &
\frac{p}{q} \frac{\sqrt{1-\theta^2}}{2} \Delta \mu
\\*[2mm]
\frac{q}{p} \frac{\sqrt{1-\theta^2}}{2} \Delta \mu \ &
\mu + \frac{\Delta \mu}{2} \theta
\end{array}
\right)
+
\left(
\begin{array}{cc}
\chi & 0\\
0 & \bar \chi
\end{array}
\right).
\label{master-matter}
\end{equation}
A few features are worth mentioning.
Firstly,
$H^\prime_{12}= H_{12}$ and $H^\prime_{21}= H_{21}$.
As a result,
$q^\prime/p^\prime=q/p$.
In particular,
the CP- and T-violating parameter $\delta$,
which depends on $|q^\prime/p^\prime|=|q/p|$,
is the same in vacuum and in the presence of matter.
Therefore,
the parameters in vacuum and in matter are related through,
\begin{eqnarray}
\mu^\prime &=& \mu + \frac{\chi + \bar \chi}{2},
\nonumber\\
\Delta \mu^\prime &=& \sqrt{(\Delta \mu)^2 + 
2 \theta\, \Delta \mu\, \Delta \chi + (\Delta \chi)^2}
= \Delta \mu\, \sqrt{1 + 4 r\, \theta + 4 r^2},
\nonumber\\
\theta^\prime &=&
\frac{\Delta \mu\, \theta + \Delta \chi}{
\sqrt{(\Delta \mu)^2 + 2 \theta\, \Delta \mu\, \Delta \chi + 
(\Delta \chi)^2}}
= \frac{\theta + 2 r}{\sqrt{1 + 4 r\, \theta + 4 r^2}},
\label{matter-vacuum-general}
\end{eqnarray}
where $\Delta \chi = \bar \chi - \chi$,
and we have introduced the `regeneration parameter'
$r=\Delta \chi/(2 \Delta \mu)$.
It will also prove convenient to find
\begin{equation}
\sqrt{1-\theta^{\prime 2}}
=
\frac{\Delta \mu\, \sqrt{1-\theta^2}}{\sqrt{(\Delta \mu)^2 
+ 2 \theta\, \Delta \mu\, \Delta \chi + (\Delta \chi)^2}}
=
\sqrt{\frac{1-\theta^2}{1 + 4 r\, \theta + 4 r^2}}.
\label{also-sin}
\end{equation}
Secondly,
it is clear from Eq.~(\ref{master-matter}),
and also from Eqs.~(\ref{matter-vacuum-general}) and
(\ref{also-sin}),
that the flavor-diagonal matter effects considered here
act just like violations of CPT.
Thirdly,
we expect the matter effects to be much larger than any
(necessarily small) CPT-violation that there might be already
present in vacuum.
Therefore,
we may take $\theta=0$ to get
\begin{eqnarray}
\mu^\prime &=& \mu + \frac{\chi + \bar \chi}{2},
\nonumber\\
\Delta \mu^\prime &=& \sqrt{(\Delta \mu)^2 + (\bar \chi - \chi)^2}
= \Delta \mu\, \sqrt{1 + 4 r^2},
\nonumber\\
\theta^\prime &=&
\frac{\bar \chi - \chi}{\sqrt{(\Delta \mu)^2 + (\bar \chi - \chi)^2}}
= \frac{2 r}{\sqrt{1 + 4 r^2}},
\label{matter-vacuum-particular}
\end{eqnarray}
and $\sqrt{1 - \theta^{\prime 2}}=1/\sqrt{1 + 4r^2}$.
We stress that Eq.~(\ref{master-matter}) is completely general,
as will be the time evolution based on it.

The time evolution in matter is now trivial to find.
It is given in Eqs.~(\ref{evolution-in-general}) 
[or, alternatively, in Eqs.~(\ref{usual-time-evolution})]
and (\ref{g+-}),
with the unprimed quantities substituted by the primed quantities.
This solution had been found for the kaon system by Good \cite{Goo57},
building on earlier work by Case \cite{Cas56}.
Recent re-derivations may be found in references \cite{Fet96} and \cite{Sch98}.
In all these articles,
the authors write a new evolution equation obtained by combining
the diagonalized form of $\mbox{\boldmath $H$}$ with the new term 
$\mbox{\boldmath $H$}_{\rm nuc}$
written in the $\{K_L, K_S\}$ basis.
Thus,
they would seem to be solving a new complicated set of equations:
the so-called `Good equations'.
In the method presented here,
we have made no reference to `new' differential equations.
We had already solved the most general evolution equation once and for all,
Eqs.~(\ref{usual-time-evolution});
and we had seen how $\mbox{\boldmath $H$}$ 
could be written in terms of observables,
Eq.~(\ref{master}).
All we had to do was to refer back to those results.

It should also be pointed out that this matrix formulation is
very useful whenever we have non-uniform materials.
For example,
one might wish to study an experiment in which a kaon
beam traverses vacuum, matter, and then vacuum again before
it decays.
Or a beam that traverses copper, carbon and then tungsten.
In the matrix formulation,
all we have to do is multiply three evolution matrices
\begin{equation}
\exp{\left[ -i {\cal H} t_1 \right]}\,
\exp{\left[ -i {\cal H} (t_2 - t_1) \right]}\,
\exp{\left[ -i {\cal H} (t_3 - t_2) \right]},
\end{equation}
each given by Eq.~(\ref{evolution-in-general}).

\section{On the (mathematical) relation with neutrino oscillations}
\label{sec:boost}

\subsection{Boosted frames}

As we have mentioned before,
the evolution equation (\ref{Schrodinger-equation}) in which
our study is based has been written in the rest frame of the
$P^0 - \overline{P^0}$ system.
We denote this explicitly by
\begin{equation}
i \frac{d}{dt_{\rm rest}} |\psi(t_{\rm rest})\rangle
= \mbox{\boldmath $H$} |\psi(t_{rest})\rangle.
\label{equation-rest}
\end{equation}
The advantage of doing this is that in the rest frame the
energy is given simply by $E=m$.
As a result,
the time parameter which appears in the solutions
presented in Eqs.~(\ref{usual-time-evolution}) 
or (\ref{usual-time-evolution-BS}),
through the time dependent functions $g_{\pm}(t)$ defined in Eq.~(\ref{g+-}),
is really $t_{\rm rest}$.

Now imagine that we wished to have Eq.~(\ref{equation-rest})
given in a boosted frame (named the lab frame from now on).
In that case we would start by noticing that both the energy and
the time are altered in the boosted frame. They become
\begin{eqnarray}
E_{\rm lab} &=& m\, \gamma,
\nonumber\\
t_{\rm lab} &=& \gamma\, t_{\rm rest} 
= \frac{E_{\rm lab}}{m}\, t_{\rm rest}.
\end{eqnarray}
Ignoring the matrix structure for the time being,
Eq.~(\ref{equation-rest}) would change schematically into
\begin{equation}
i \frac{d}{dt_{\rm lab}} |\psi(t_{\rm lab})\rangle
= m\, \gamma\,  |\psi(t_{\rm lab})\rangle
= E_{\rm lab} |\psi(t_{\rm lab})\rangle,
\label{equation-lab}
\end{equation}
as it had to.
Now,
if the boost is much larger that the mass,
$p \gg m$, we may use
\begin{equation}
E_{\rm lab} = \sqrt{p^2+m^2}
\sim p + \frac{m^2}{2p} + \cdots
\sim p + \frac{m^2}{2E} + \cdots.
\end{equation}

However, we do not need to do this.
We have already found the solution to Eq.~(\ref{equation-rest})
in the rest frame.
In order to change it into the lab frame all we have to do is to
substitute $t_{\rm rest}$ in the time evolution functions of 
Eq.~(\ref{g+-}) by $t_{\rm rest} = t_{\rm lab}/\gamma$.
We notice that $1/\gamma=m/E$.
Therefore,
when written in terms of $t_{\rm lab}$ the time evolution
functions of Eq.~(\ref{g+-}) become
\begin{equation}
g_{\pm}(t_{\rm rest})
=
e^{-i m^2 t_{\rm lab}/E}\, e^{- \Gamma\, m\, t_{\rm lab}/(2 E)}
\left\{
\begin{array}{c}
\cos{(\frac{m}{E} \frac{\Delta \mu}{2}\, t_{\rm lab})}\\*[2mm]
-i \sin{(\frac{ m}{E} \frac{\Delta \mu}{2}\, t_{\rm lab})}
\end{array}
\right.
\label{g+--boosted}
\end{equation}
And,
using $\Delta m = m_a - m_b$ and $m = (m_a + m_b)/2$,
we realize that the argument of the trigonometric functions is
given by
\begin{equation}
\frac{m}{E}\, \frac{\Delta \mu}{2}\, t_{\rm lab}
= \frac{m_a^2 - m_b^2}{4 E}\,  t_{\rm lab}
- \frac{i}{4} \frac{\Delta \Gamma\, m}{E}\,  t_{\rm lab}.
\label{argument-boosted}
\end{equation}

\subsection{A neutrino-like oscillation}

For the comparison with neutrinos,
it is most convenient to use the parametrization of
the CP-violating quantities discussed in the first subsection
of appendix~\ref{app:others}.
To obtain relations that mimic those in the neutrino system,
we compute the probability that $P^0$ becomes
$\overline{P^0}$ using Eqs.~(\ref{usual-time-evolution-BS}),
(\ref{g+--boosted}),
(\ref{argument-boosted}),
setting $\mbox{Im}\, \phi_R = \mbox{Im}\, \theta_R=0$,
and letting $\Gamma = \Delta \Gamma \sim 0$
(another way of thinking about this limit is to suppose
that we are performing an experiment in a time scale much
smaller than the mesons' decay time).
We find
\begin{equation}
{\left| \langle \overline{P^0}| P^0 (t_{\rm lab}) \rangle \right|}^2
=
\sin^2{\theta_R}\,
\sin^2{\left( \frac{m_a^2 - m_b^2}{4 E}\,  t_{\rm lab} \right)}.
\label{with-sin-2}
\end{equation}
If, instead,
the experiment is performed in matter,
we obtain
\begin{eqnarray}
{\left| \langle \overline{P^0}| P^0 (t_{\rm lab}) \rangle \right|}^2
&=&
\left| \sin^2{\theta_R^\prime} \right|\,
\sin^2{\left( \frac{m_a^2 - m_b^2}{4 E}\,  t_{\rm lab} \right)},
\nonumber\\*[3mm]
&=&
\frac{\sin^2{\theta_R}}{\left| 1 - 4 r\, \cos{\theta_R} + 4 r^2 \right|}
\sin^2{\left( \frac{m_a^2 - m_b^2}{4 E}\,  t_{\rm lab} \right)}
\nonumber\\*[3mm]
&=&
\frac{\sin^2{\theta_R}}{
\left| (\cos{\theta_R} - 2 r)^2 + \sin^2{\theta_R} \right|}
\sin^2{\left( \frac{m_a^2 - m_b^2}{4 E}\,  t_{\rm lab} \right)},
\label{neutrino-mix}
\end{eqnarray}
where we have used Eq.~(\ref{relate-vacuum-matter-BS}) in getting 
to the second line,
and 
\begin{equation}
r = \frac{\Delta \chi}{2 \Delta \lambda}
= \frac{\Delta \chi}{2 \Delta m}
= \frac{\Delta \chi/\gamma}{2 \Delta m/\gamma}
= \frac{E}{m_a^2-m_b^2} \frac{\Delta \chi}{\gamma}.
\end{equation}
Eq.~(\ref{neutrino-mix}) exhibits a resonance structure because the time
independent coefficient reaches its maximum if 
$2 r = \cos{\theta_R}$.
For the final step in the connection to neutrinos,
we look at this case further by assuming that the imaginary
part of $r$ ($\Delta \chi$),
which is proportional to $\sigma_{\rm tot}$ in appendix~\ref{app:A},
is negligible.
Then $\Delta \chi$ is real and we may parametrize
\begin{equation}
V
\equiv
\frac{\mbox{Re} \Delta \chi}{ \gamma}.
\end{equation}
%
%
As a result,
\begin{equation}
r = \frac{E\, V}{m_a^2-m_b^2}
\end{equation}
is real and the resonance condition,
which becomes
\begin{equation}
\frac{2 E\, V}{m_a^2-m_b^2} = \cos{\theta_R},
\label{neutrino-resonance}
\end{equation}
can be satisfied for $\theta_R$ real.
Eq.~(\ref{neutrino-mix}) and the resonance condition in 
Eq.~(\ref{neutrino-resonance}) are in exactly the same form as
the usual discussions of neutrino oscillations in matter \cite{wavepackets}.

Although there is this mathematical connection between neutrino
oscillations and $P^0 - \overline{P^0}$ oscillations,
the situations are physically very different.
Indeed,
it is important to notice that there are no CPT relations between
the two neutrino species involved in neutrino oscillation,
and the vacuum mixing angle $\theta_R$ in Eq.~(\ref{with-sin-2}) may be
small.
Eq.~(\ref{neutrino-mix}) shows that,
even if $\theta_R$ is small,
the effective mixing angle in matter will be large when one
hits the resonance condition in Eq.~(\ref{neutrino-resonance}).
In contrast,
as we show in appendix~\ref{app:others},
in the $P^0 - \overline{P^0}$ system,
the deviation of $\cos{\theta_R}$ from zero measures violations of CPT.
Assuming CPT conservation in the $P^0 - \overline{P^0}$
system,
$\sin{\theta_R} = 1$ and the vacuum transition in Eq.~(\ref{with-sin-2})
already reaches unity (at select times)\footnote{Recall that we have
assumed $\Gamma$=0 and $\mbox{Im}\, \phi_R = 0$ (T conservation).
When $\Gamma \neq 0$, the right hand side of  Eq.~(\ref{with-sin-2}) 
appears multiplied by $\exp{(- \Gamma t_{\rm rest})}$.}.
Said otherwise,
the small mixing angle mixing discussed in neutrino oscillations
in vacuum,
is (in the connection presented here) the mathematical analogue
of large violations of CPT in the $P^0 - \overline{P^0}$ system.

\section{Conclusions}
\label{sec:conclusions}

We have shown that the presence of T violation in the
neutral meson systems implies that the corresponding
effective Hamiltonian $\mbox{\boldmath $H$}$ does not commute with itself.
Therefore,
$\mbox{\boldmath $H$}$ cannot be diagonalized by an unitary 
transformation and we
must introduce the reciprocal basis.
This basis must be used in order to obtain the correct
form for some physical observables,
such as the parameter $\lambda$ in the decays $B_d \rightarrow J/\psi K_S$.
But,
working with the reciprocal basis is a blessing
rather than a nuisance.
We show that using the reciprocal basis has the following advantages:
\begin{itemize}
\item the relation between the effective Hamiltonian matrix when
written in the mass and flavor basis is simply obtained and
easily inverted, thus providing a parametrization of $\mbox{\boldmath $H$}$
in terms of measurable quantities;
\item one obtains a one line derivation of the evolution of the states;
\item propagation in matter is reduced to the case of propagation
in vacuum,
with the vacuum and matter parameters related in a trivial fashion,
without any recourse to the Good equations;
\item the propagation in non-uniform media is reduced to a multiplication
of evolution matrices.
\end{itemize}
It is true that some of these results can be obtained without
using the reciprocal basis.
But,
as we have tried to illustrate in the article,
this concept is not only needed but, when used, greatly simplifies
the various derivations.
In addition,
we can use this formalism to highlight the similarity between the
matter effects in the $P^0 - \overline{P^0}$ systems and 
the matter effects in neutrino oscillations.

\acknowledgments

I would like to thank A.\ Barroso, L.\ Lavoura, Y.\ Nir, 
H.\ R.\ Quinn, M.\ Weinstein, and L.\ Wolfenstein for reading this
manuscript and for their useful suggestions.
I am extremely indebted to A.\ Barroso, L.\ Lavoura,
and  J.\ A.\ Perdig\~{a}o Dias da Silva for many discussions on this topic. 
This work is supported in part by the Department of Energy 
under contract DE-AC03-76SF00515.
The work of J.\ P.\ S.\ is supported in part by Fulbright,
Instituto Cam\~oes, and by the Portuguese FCT, under grant
PRAXIS XXI/BPD/20129/99	and contract CERN/S/FIS/1214/98.

\appendix


\section{Matter effects in the rest and laboratory frames}
\label{app:A}

In this appendix we show how $\mbox{\boldmath $H$}_{\rm nuc}$ is related to
physical cross-sections and what is the form of this
relation in the rest and laboratory frames.
This is relevant for Eq.~(\ref{nuc}) and for those wishing
to expand on the analogy with the neutrino oscillations discussed
at the end of section \ref{sec:boost}.
We follow here the notation of references \cite{Gas74} and \cite{Sch98}.

Let us consider the evolution of a coherent wave-packet $\phi$ 
with wavenumber $k$ in the laboratory frame,
\begin{equation}
\frac{d}{dz} \phi = i k\, \phi.
\end{equation}
In the presence of a block of material at rest in the laboratory
frame,
the wavenumber suffers a shift given approximately by
\begin{equation}
k^\prime - k \approx \frac{2 \pi N}{k} f(0)
= N \left(
\frac{2 \pi}{k} \mbox{Re} f(0)
+
i \frac{\sigma_{\rm tot}}{2}
\right),
\end{equation}
where $N$ is the density of scattering centers in the medium and
$f(0)$ is the elastic forward scattering amplitude.
On the last equality,
we have used the fact that the imaginary part of $f(0)$ is related  to the 
total cross section $\sigma_{\rm tot}$ by the optical theorem,
\begin{equation}
\mbox{Im} f(0) = \frac{k}{4 \pi} \sigma_{\rm tot}.
\end{equation}
We also recall that
\begin{equation}
\left| f(\theta)\right|^2 = \frac{1}{2 \pi} \frac{d \sigma}{d \cos{\theta}}.
\label{def-cross-section}
\end{equation}
In this equation (and only here),
$\theta$ refers to the scattering angle in the laboratory frame.

We conclude that
the presence of matter changes the evolution in vacuum by an amount
\begin{equation}
i \frac{d}{d t_{\rm lab}} \phi
= 
- v (k^\prime - k) \phi
=
- \frac{2 \pi N}{k/v} f(0)\, \phi,
\end{equation}
where $v$ is the beam velocity in the lab frame and $z = v\, t_{\rm lab}$.
To change into the rest frame of the beam we notice that
$t_{\rm lab} = \gamma\,  t_{\rm rest}$ and $k = m \gamma v $,
where $\gamma = 1/\sqrt{k^2 + m^2}$,
leading to
\begin{equation}
i \frac{d}{d t_{\rm rest}} \phi
= - \frac{2 \pi N}{m} f(0)\, \phi.
\end{equation}

When studying the $P^0 - \overline{P^0}$ systems,
we denote by $f$ ($\bar f$) the elastic forward scattering amplitude of
$P^0$ ($\overline{P^0}$).
Therefore,
the new contribution in the $P^0 - \overline{P^0}$ rest frame
is given by
\begin{equation}
i \frac{d}{d t_{\rm rest}} | \psi(t_{\rm rest}) \rangle
=
\left(
\begin{array}{cc}
\chi & 0\\
0 & \bar \chi
\end{array}
\right)
| \psi(t_{\rm rest}) \rangle,
\end{equation}
where \cite{Sch98}
\begin{equation}
\chi = - \frac{2 \pi N}{m} f\ \ \ \mbox{and}\ \ \
\bar \chi = - \frac{2 \pi N}{m} \bar f,
\end{equation}
leading to Eq.~(\ref{nuc}).


\section{Other parametrizations for T and CPT violation}
\label{app:others}

The way we parametrize T and CPT violation in the mixing of neutral mesons
is different from the parametrizations used by some other authors.
For ease of reference,
we collect here formulae summarizing the relationships
among different parametrizations.

\subsection{The parameters $\phi_R$ and $\theta_R$}

Some authors (for instance \cite{Kobayashi92})
introduce two complex angles $\theta_R$ and $\phi_R$ by writing
\begin{eqnarray}
p_a = N_a \cos{\frac{\theta_R}{2}}, \ \ \ \ \ 
q_a = N_a e^{i \phi_R} \sin{\frac{\theta_R}{2}},
\nonumber\\
p_b = N_b \sin{\frac{\theta_R}{2}}, \ \ \ \ \ 
q_b = N_b e^{i \phi_R} \cos{\frac{\theta_R}{2}}.
\end{eqnarray}
Then,
\begin{eqnarray}
\frac{q}{p} &=& e^{i \phi_R},
\nonumber\\*[2mm]
\delta &=& \tanh{\left( \mbox{Im}\, \phi_R \right)},
\nonumber\\*[2mm]
\theta &=& - \cos \theta_R,
\end{eqnarray}
and $\sqrt{1 - \theta^2} = \sin \theta_R$.
CPT is violated if and only if $\cos{\theta_R} \neq 0$.
T is violated if and only if $\mbox{Im}\, \phi_R \neq 0$.
Some authors use a particular phase convention
and claim that $\mbox{Re}\, \phi_R \neq 0$ also corresponds to T violation.
Clearly this statement is false since the phase of $q/p$ has
no physical meaning;
we know that there is one and only one T- and CP-violating quantity in 
$\mbox{\boldmath $H$}$.

With this notation,
Eqs.~(\ref{X-paramet}),
(\ref{X-1-paramet}),
(\ref{master}),
(\ref{evolution-in-general}),
and (\ref{usual-time-evolution})
become
\begin{equation}
\mbox{\boldmath $X$} =
\left(
\begin{array}{cc}
\cos{\frac{\theta_R}{2}}\ &  - \sin{\frac{\theta_R}{2}}\\*[2mm]
e^{i \phi_R} \sin{\frac{\theta_R}{2}}\ & e^{i \phi_R} \cos{\frac{\theta_R}{2}} 
\end{array}
\right)
\left(
\begin{array}{cc}
N_a & 0\\*[2mm]
0 & - N_b
\end{array}
\right),
\label{X-paramet-BS}
\end{equation}
\begin{equation}
\mbox{\boldmath $X$}^{-1} =
\left(
\begin{array}{cc}
N_a^{-1} & 0\\*[2mm]
0 & - N_b^{-1}
\end{array}
\right)
\left(
\begin{array}{cc}
\cos{\frac{\theta_R}{2}}\ & e^{- i \phi_R} \sin{\frac{\theta_R}{2}}\\*[2mm]
- \sin{\frac{\theta_R}{2}}\ & e^{-i \phi_R} \cos{\frac{\theta_R}{2}} 
\end{array}
\right),
\label{X-1-paramet-BS}
\end{equation}
\begin{equation}
\mbox{\boldmath $H$}
=
\left(
\begin{array}{cc}
\mu + \cos{\theta_R}\, \frac{\Delta \mu}{2} \ &
e^{-i \phi_R} \sin{\theta_R}\, \frac{\Delta \mu}{2}
\\*[2mm]
e^{i \phi_R} \sin{\theta_R}\, \frac{\Delta \mu}{2} \ &
\mu - \cos{\theta_R}\, \frac{\Delta \mu}{2}
\end{array}
\right),
\label{master-BS}
\end{equation}
\begin{equation}
\exp{(- i {\cal H} t)}
=
\left(
\begin{array}{cc}
| P^0 \rangle, & | \overline{P^0} \rangle
\end{array}
\right)
\left(
\begin{array}{cc}
g_+(t) + \cos{\theta_R}\, g_-(t)\  & 
	e^{- i \phi_R} \sin{\theta_R}\, g_-(t)\\*[2mm]
e^{i \phi_R} \sin{\theta_R}\, g_-(t)\  & 
	g_+(t) - \cos{\theta_R}\, g_-(t)
\end{array}
\right)
\left(
\begin{array}{c}
\langle P^0 | \\
\langle \overline{P^0} |
\end{array}
\right),
\label{evolution-in-general-BS}
\end{equation}
and
\begin{eqnarray}
| P^0(t) \rangle
&=&
\left[ g_+(t) + \cos{\theta_R}\, g_-(t) \right] | P^0 \rangle
+
e^{i \phi_R} \sin{\theta_R}\, g_-(t)\, | \overline{P^0} \rangle,
\nonumber\\*[3mm]
| \overline{P^0}(t) \rangle
&=&
e^{- i \phi_R} \sin{\theta_R}\, g_-(t) | P^0 \rangle
+
\left[ g_+(t) - \cos{\theta_R} \, g_-(t) \right] | \overline{P^0} \rangle,
\label{usual-time-evolution-BS}
\end{eqnarray}
respectively.

Finally,
the relation between the matter and vacuum parameters described
in Eqs.~(\ref{matter-vacuum-general}) and (\ref{also-sin})
become
\begin{eqnarray}
\mu^\prime &=& \mu + \frac{\chi + \bar \chi}{2},
\nonumber\\
\Delta \mu^\prime &=& \sqrt{(\Delta \mu)^2 -
2 \cos{\theta_R}\, \Delta \mu\, \Delta \chi + (\Delta \chi)^2}
= \Delta \mu\, \sqrt{1 - 4 r\, \cos{\theta_R} + 4 r^2},
\nonumber\\
\cos{\theta_R^\prime} &=&
\frac{\Delta \mu\, \cos{\theta_R} - \Delta \chi}{
\sqrt{(\Delta \mu)^2 - 2 \cos{\theta_R}\, \Delta \mu\, \Delta \chi + 
(\Delta \chi)^2}}
= \frac{\cos{\theta_R} - 2 r}{\sqrt{1 - 4 r\, \cos{\theta_R} + 4 r^2}},
\nonumber\\
\sin{\theta_R^\prime}
&=&
\frac{\Delta \mu\, \sin{\theta_R}}{\sqrt{(\Delta \mu)^2 
- 2 \cos{\theta_R}\, \Delta \mu\, \Delta \chi + (\Delta \chi)^2}}
=
\frac{\sin{\theta_R}}{\sqrt{1 - 4 r\, \cos{\theta_R} + 4 r^2}},
\label{relate-vacuum-matter-BS}
\end{eqnarray}
and $\phi_R^\prime=\phi_R$.

\subsection{The parameters $\epsilon_S$ and $\delta_S$}

Other authors (for instance \cite{CPLEAR-CPT,Briere89})
use two complex parameters,
$\epsilon_S$ and $\delta_S$,
and write
\begin{eqnarray}
\frac{q_a}{p_a} &=& \frac{1 - \epsilon_S + \delta_S}
{1 + \epsilon_S - \delta_S},
\nonumber \\
\frac{q_b}{p_b} &=& \frac{1 - \epsilon_S - \delta_S}
{1 + \epsilon_S + \delta_S}.
\end{eqnarray}
Obviously then,
\begin{eqnarray}
\frac{q}{p} &=& 
\sqrt{\frac{\left( 1 - \epsilon_S \right)^2 - \delta_S^2}
{\left( 1 + \epsilon_S \right)^2 - \delta_S^2}},
\nonumber\\*[2mm]
\delta &=&
\frac{8 \mbox{Re} \left[ \epsilon_S^\ast
\left( 1 + \epsilon_S^2 - \delta_S^2 \right) \right]}{
\left( \left| \left( 1 + \epsilon_S \right)^2 - \delta_S^2 \right| + 
\left| \left( 1 - \epsilon_S \right)^2 - \delta_S^2 \right|
\right)^2},
\nonumber\\*[2mm]
\theta &=& 
\frac{2 \delta_S}{1 + \delta_S^2 - \epsilon_S^2}.
\end{eqnarray}
CPT invariance corresponds to $\delta_S = 0$.
T invariance corresponds to $\mbox{Re} \left[ \epsilon_S^\ast
\left( 1 + \epsilon_S^2 - \delta_S^2 \right) \right] = 0$.
The authors who use this parametrization,
however,
always do so in conjunction with the assumption
that $\delta_S$ and $\epsilon_S$ are small.
Then,
\begin{eqnarray}
\delta &\approx& 2\, \mbox{Re}\, \epsilon_S,
\nonumber\\
\theta &\approx& 2 \delta_S.
\end{eqnarray}
Moreover,
$\sqrt{1 - \theta^2} \approx 1 - 2 \delta_S^2$.

It should be kept in mind that the $R$-parametrization is exact and general,
while the $S$-parametrization is interesting only when using
a phase convention
${\cal C P} | P^0 \rangle = \pm | \overline{P^0} \rangle$,
which implies that CP conservation corresponds to vanishing
$\delta_S$ and $\epsilon_S$.


\begin{references}
\bibitem{before}
As far as we know,
a similar expression was first derived in reference \cite{Alv99}.
The form presented here highlights the T- and CPT-violating quantities,
and will now be used in new contexts.
\end{references}
\end{document}